\documentclass[twocolumn, preprintnumbers,amsmath,amssymb]{revtex4}
\usepackage{graphicx}
\usepackage{dcolumn}
\usepackage{bm}

\newcommand{\be}{\begin{equation}}
\newcommand{\ee}{\end{equation}}

\begin{document}

\title{A dynamic jamming point for shear thickening suspensions}
\author{Eric Brown}
%\email{embrown@uchicago.edu}
\author{Heinrich M. Jaeger}
\affiliation{The James Franck Institute, University of Chicago, Chicago, IL 60637}
\date{\today}

\begin{abstract} We report on rheometry measurements to characterize critical behavior in two model shear thickening suspensions: cornstarch in water and glass spheres in oil.  The slope of the shear thickening part of the viscosity curve is found to increase dramatically with packing fraction and diverge at a critical packing fraction $\phi_c$. The magnitude of the viscosity and the yield stress are also found to have scalings that diverge at $\phi_c$.  We observe shear thickening as long as the yield stress is less than the stress at the viscosity maximum.  Above this point the suspensions transition to purely shear thinning.  Based on these data we present a dynamic jamming phase diagram for suspensions and show that a limiting case of shear thickening corresponds to a jammed state.  

\end{abstract}

%\pacs{83.85.Vb, 83.80.Hj, 83.60.Rs}

\maketitle

In Newtonian fluids the viscosity does not change with applied shear rate, while non-Newtonian fluids usually show a decrease of viscosity when sheared faster, i.~e., they shear thin. The opposite behavior, shear thickening, is less common but can be quite dramatic: beyond a certain shear rate the viscosity increases potentially by orders of magnitude.  This behavior is reversible so the stress relaxes when the shear is removed.  Reported shear thickening fluids are usually densely packed colloids or suspensions \cite{Ba89, MW01a, HL05}. Strong shear thickening can be observed for example with cornstarch in water.  Shear thickening is a concern across a range of industrial processes \cite{Ba89, LHLD04} and is of interest for the ability to absorb energy from impacts \cite{LWW03}.

It has been suggested \cite{HAC01, BBS02, LDH03, HFC03, MB04b, HCFS05, CHH05,DP05, HL05, LDHH05, SK05,ILSH06,FHBOB08} that shear thickening is related to the phenomenon of jamming. %whereby an amorphous, liquid-like particle aggregate  under confinement develops rigidity and ceases to flow
The concept of a jamming transition, however, applies to the limit of vanishing shear rate and is exemplified by the onset of glassy behavior with a seemingly diverging viscosity in molecular liquids, dense packings of colloids, or macroscopic granular materials, leading to the appearance of a yield stress below which there is no flow \cite{LN98, LN01, OSLN03}.  Shear thickening, on the other hand, occurs at non-zero shear rate. While a yield stress has been measured in some shear thickening fluids \cite{LDH03}, no linkage has been established between such yield stress and the observed shear thickening behavior.  Dense shear thickening fluids have been reported to exhibit seemingly discontinuous jumps in stress with increasing shear rate \cite{Hoffmann, Laun94, BW96, FH96, BBS02, EW05, LDHH05, FHBOB08}. However, this discontinuity has not yet been characterized quantitatively.  %This gives the impression that the material becomes solid-like at a non-zero shear rate.  Structurally, shear thickening has been associated with the formation of particle clusters \cite{BB88, MW01a, MB04a, EW05}, 
Phenomenological shear thickening models have shown stress distributions that are similar to those of force chains in jammed systems \cite{HAC01, MB04b}.  However, the mild shear thickening behavior found in simulations \cite{BB85, MVB96, MB04b, GCNC08} so far cannot reproduce the dramatic increases in viscosity with shear rate observed in many experiments.     As a result,  the connection between shear thickening and the onset of jamming has been qualitative and many details of the relationship remain unresolved.

% and has been associated with flow-induced formation of stress-bearing particle arrangements.  As a consequence of hydrodynamic interactions, shear thickening can be observed already at low packing fractions, and for most colloid concentrations behaves differently from jamming in the sense that the viscosity may be large but does not diverge.  [speculation]

% While several simulations have shown mild shear thickening behavior \cite{BB85, MVB96, HAC01, GCNC08} none of them have captured this dramatic shear thickening behavior.  Phenomenological shear thickening models have shown stress distributions that are similar to those of force chains in jammed systems \cite{HAC01}.  Bertrand et.~al.\cite{BBS02} reported a {\it shear-induced jamming} that is irreversible under shear but that has not been found to generally occur in shear thickening suspensions and we restrict our study to reversible shear thickening.  Despite these references, shear thickening suspensions are not jammed because they still flow and lack rigidity.  

Here we characterize the stress-shear rate discontinuity and the relationship to the yield stress through rheological measurements.  This includes the first systematic characterization of shear thickening rheology as a function of packing fraction for suspensions of non-Brownian particles.  We find that discontinuous shear thickening is a limiting behavior which is approached at a critical packing fraction where the onset shear rate of shear thickening approaches zero and the yield stress jumps dramatically.  In other words, the limiting case of discontinuous shear thickening corresponds to a jammed state.  We then develop a phase diagram to delineate the shear thickening and jammed regimes.  

%\section{Experimental methods}

We present below experimental results from two rather different hard particle suspensions, cornstarch in water and  glass spheres in mineral oil, to demonstrate the generality of our results. Cornstarch (average particle diameter of 14 $\mu$m)  was chosen as a prototypical shear thickener, while glass spheres (88-125 $\mu$m diameter) have the advantages of better defined particle properties and better studied packing properties.  Measurements were performed in a rheometer using Couette (inner rotating cylinder of diameter $26.6$ mm, gap of $1.13$ mm)  or parallel plate ($25$ mm diameter rotating top plate) geometries.   All data presented were taken with a constant applied torque converted to a shear stress $\tau$ (based on a Newtonian flow profile).   The shear rate $\dot\gamma$ is defined by the measured rotation velocity over gap size.  We define viscosity by $\eta\equiv\tau/\dot\gamma$. Samples were pre-sheared before experiments and measurements were performed with increasing as well as decreasing stress ramps to check for repeatability.  Reported packing fractions $\phi$ are based on measured particle and fluid masses mixed together before shearing.   %Absolute errors on the packing fraction are around 0.005 from the density calibration while relative errors can be smaller.

The starch particles were suspended in water, density-matched to 1.59 g/mL by dissolving CsCl.  From measurements with optical tweezers we found no significant particle interactions in the stress range of the experiments.  We measured packing fractions using the weight of starch at ambient conditions of 23$^{\circ}$C and 42\% humidity which includes some water weight.   Glass spheres with hydrophobic coating were used to optimize dispersion in mineral oil (viscosity 58 mPa$\cdot$s) to minimize the yield stress.  Since glass spheres are denser than mineral oil,  measurements were performed in the parallel plate geometry with a gap size of 0.5 mm. This minimized pressure on the packing due to its own weight.   Measurements reported here are done with no oil extending outside the parallel plates so that the particles are confined to the space between the plates by surface tension. %Extension of the fluid outside the plates is known to increase the critical shear rate \cite{FHBOB08}.  

%\section{Results}

\begin{figure}
\includegraphics[width=3.4in]{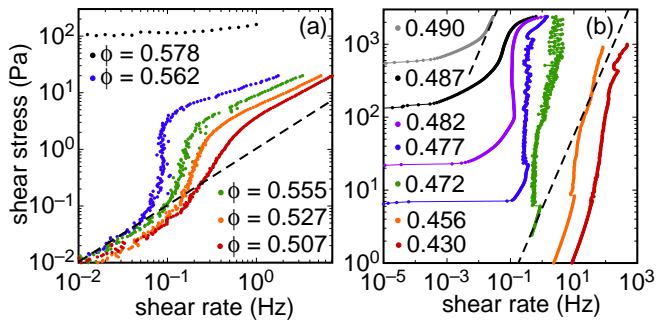}
\caption{Shear stress $\tau$ vs.~shear rate $\dot\gamma$ for packing fractions $\phi$, as indicated (color online).  (a) Glass spheres in mineral oil. (b) Cornstarch in water.  Dashed lines:  slope 1 indicating fixed viscosity. }
\label{fig:stressvsrate}
\end{figure}

Figure \ref{fig:stressvsrate} gives the evolution of stress vs.~shear rate curves as the packing fraction is increased.   On a log-log plot as shown, a slope of 1 corresponds to Newtonian flow, indicated by dashed lines for reference, a slope between zero and unity corresponds to shear thinning, while a slope greater than unity signals shear thickening. The overall steepness of the traces within the shear thickening region is seen to increase with $\phi$ and to approach a vertical line where shear thickening becomes discontinuous.  Another feature is that shear thickening is associated with a certain intermediate stress range that varies little with packing fraction.  For stresses either larger than the upper limit of the shear thickening region or smaller than the shear thickening onset, shear thinning behavior is observed.  At lower $\phi$ the slopes gradually approach 1 at all stress ranges so there is a gradual transition to Newtonian flow.  The behavior described above is similar to what has been found by careful measurements in shear thickening colloids \cite{EW05}.  The similarity is notable because it is usually assumed that Brownian motion and electrostatics are important factors in shear thickening \cite{MW01a} which are insignificant for our larger particles.  At sufficiently large $\phi$,  the curves exhibit a non-zero stress value in the limit of zero shear rate, i.~e., a yield stress. Given our stress resolution around $10^{-2}$ Pa in our rheometer, this is most clearly seen in the cornstarch data where the yield stress is larger.  The yield stress is seen to encroach on the shear thickening stress range at high packing fractions above which there is only shear thinning and no shear thickening. %Shear thickening behavior therefore persists only up to a certain maximum yield stress.  In samples with yield stress larger than this maximum, the suspension will transition to purely shear thinning behavior.  

\begin{figure}                                                
\centerline{\includegraphics[width=3in]{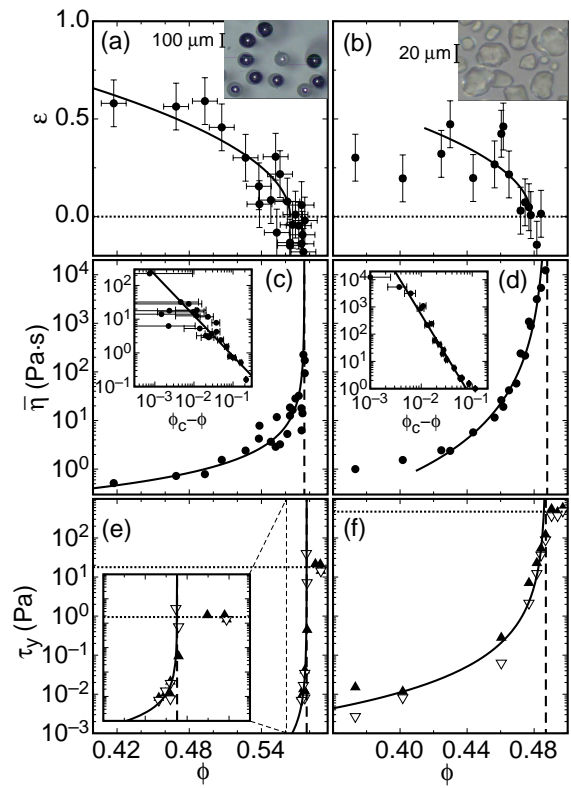}}
\caption{ Evolution of viscosity and yield stress with packing fraction $\phi$ for glass spheres in mineral oil (a,c,e) and cornstarch in water (b,d,f). (a,b) Inverse logarithmic slope $\epsilon$ of stress-shear rate traces in the shear thickening regime, defined by  $\tau\propto\dot\gamma^{1/\epsilon}$.  Insets show micrographs of the particles. (c,d) Viscosity scale $\bar\eta$  defined as geometric mean of the viscosity in the shear thickening region.  Insets in (c,d): log-log plots of the same data relative to $\phi_c$.  (e,f) Yield stress $\tau_y$. Solid triangles:  increasing stress measurements .  Open triangles: decreasing stress measurements.  Dotted lines: the plateau yield stress $\tau_{y,max}$. Inset in (e): detail view of the region close to $\phi_c$. All panels: Solid lines are power law fits explained in text,  dashed lines are resulting values of $\phi_c$.   }
\label{fig:figcompound}                                        
\end{figure}

To quantify the stress-strain relationship in the shear thickening regions, we fit the traces locally to a power law $\eta \propto \tau^{1-\epsilon}$ which is equivalent to $\tau\propto \dot\gamma^{1/\epsilon}$ but can be fit because its slope does not diverge. The parameter $\epsilon$ depends on packing fraction and corresponds to the inverse slope of the traces in Fig.~\ref{fig:stressvsrate}. Newtonian flow corresponds to $\epsilon=1$ and a stress divergence corresponds to $\epsilon=0$.   In Figs.~2a\&b, the $\epsilon$ values plotted are from fits around the steepest portions of the stress-shear rate traces. For both starch particles and glass spheres, $\epsilon$ approaches zero at a critical packing fraction $\phi_c$ where the slope of the viscosity curve becomes divergent.    Previous qualitative descriptions of shear thickening have made a distinction between continuous and discontinuous shear thickening regimes, implying the possibility of $\epsilon=0$ over a range of $\phi$.  However, the fact that $\epsilon$ only approaches zero at $\phi_c$ suggests that discontinuous shear thickening is better thought of as a limiting behavior of shear thickening.  The value of  $\phi_c$ is obtained from a power law fit of $\epsilon\propto (\phi_c-\phi)^n$ to the data in Figs.~\ref{fig:figcompound}a\&b which gives $\phi_c=0.478\pm 0.003$ and $n=0.5\pm0.2$ for starch and $\phi_c=0.564\pm0.004$ and $n=0.5\pm0.2$ for glass with relative statistical uncertainties.  Power laws shown are fit from $\phi_c$ down to the smallest $\phi$ which is consistent with a power law fit.  When the lower end of the fit range of $\phi$ was increased, fit values of $\phi_c$ remained consistent within quoted uncertainties, however fit values for exponents varied with the fit range, so we do not claim to have measured limiting scaling exponents with certainty.  %These conclusion holds for all fits presented in this letter.

As a second indicator of a transition near $\phi_c$, we investigate the evolution of magnitude of the viscosity.  To this end, we define a characteristic magnitude $\bar\eta$ by the geometric mean of the viscosity over a fixed stress range in the shear thickening region. Figures 2c\&d show that $\bar\eta$ appears to diverge very close to the point where $\epsilon$ goes to zero.  A diverging power law $\bar\eta \propto (\phi_c-\phi)^{-n}$ is fit to the data in Figs.~\ref{fig:figcompound}c\&d  which gives $\phi_c = 0.488\pm 0.004$ and $n=3.1\pm0.3$ for starch and $\phi_c=0.576\pm0.004$ and $n=1.2\pm0.2$ for glass.  While the viscosity varies with stress the fit value of $\phi_c$ is independent of the fixed stress range chosen for the fit in the shear thickening region.  This divergent scaling is independent of the yield stress because it is in a much higher stress range, and is independent of the diverging slope because that has a mild effect on a fixed stress range. %(for starch in water we use the stress range 500 to 1000 Pa, for glass spheres in mineral oil 0.3 to 1 Pa)

The yield stress $\tau_y$ for different packing fractions is shown in Figs.~\ref{fig:figcompound}e\&f and is found to increase precipitously as the same value $\phi_c$ is approached.  Around $\phi_c$, the yield stress plateaus at a value $\tau_{y,max}$ and does not change significantly  at higher packing fractions.  A power law $\tau_y \propto (\phi_c-\phi)^{-n}$ is fit to the data in Figs.~\ref{fig:figcompound}e\&f below the plateau to obtain $\phi_c = 0.487\pm 0.003$ and $n=2.5\pm0.3$ for starch and $\phi_c=0.578\pm0.004$ for glass spheres ($n$ could not be fit with any certainty because the jump in yield stress is so dramatic).  Comparison of the packing fraction dependence of $\epsilon$, $\bar\eta$ and $\tau_y$ in Fig.~2 shows that, statistically consistent within experimental uncertainties, a single value $\phi_c$ captures the behavior of all three parameters for each of the two suspensions 

%(independently fitting power laws of the form $\propto (\phi_c-\phi)^{n}$ to the six data sets, with $n$ = 1/2 for $\epsilon$ or  $n < -1$ for $\bar\eta$ and $\tau_y$, gives $\phi_c$ values between 0.48 and 0.49 for starch, and between 0.56 and 0.58 for glass, see dashed lines). 

\begin{figure}
\includegraphics[width=2.5in]{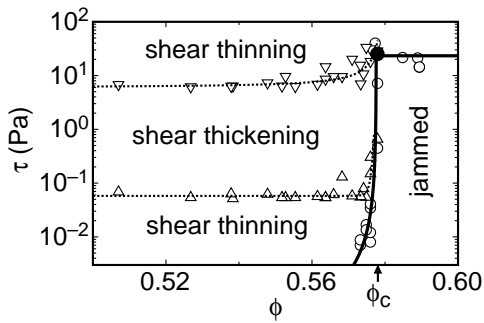}
\caption{Dynamic jamming phase diagram for shear thickening fluids. Data shown is for glass spheres in mineral oil. The solid line corresponds to the yield stress in Fig.~\ref{fig:figcompound}e, while the dotted lines indicate transitions between shear thinning and thickening.  Solid circle:  dynamic jamming point at $\phi_c$.    Open circles:  yield stress.  Up-pointing triangles: onset of shear thickening.  Down-pointing triangles:  shear thickening maximum.}
\label{fig:jam_phase}
\end{figure}

From the data in Figs.~\ref{fig:stressvsrate}\&\ref{fig:figcompound} we can assemble a phase diagram that delineates shear thinning, shear thickening and jammed regions in a parameter space given by the applied stress $\tau_y$  and the packing fraction $\phi$. Figure \ref{fig:jam_phase} shows this for the glass spheres.  The jammed region here is defined as the portion in the diagram below the yield stress.  
%The divergent scaling of the yield stress or near step function differs from usual jamming phase diagrams where the yield stress grows gradually from an onset.  
As in the usual jamming phase diagram \cite{LN98}, exceeding the yield stress leads to shear thinning flow. For $\phi < \phi_c$, a new feature in Fig.~\ref{fig:figcompound} are two boundaries that extend out from the jammed region to smaller  packing fractions. These boundaries separate shear thickening from shear thinning regimes.  Increasing stress at fixed $\phi$ will take a sample from shear thinning to thickening and back to thinning again, as seen from the slope changes of $\tau(\dot\gamma)$ in Fig.~\ref{fig:stressvsrate}.  As  $\phi_c$ is approached from below, however, the increasing yield stress pushes both boundaries upward, forcing them to approach the yield stress line that delineates the jammed region.  Once the yield stress dominates the total stress in the system  the suspension no longer exhibits shear thickening behavior but only jamming and shear thinning.  Because of the divergent scaling of the yield stress, this transition occurs at a packing fraction near to but less than $\phi_c$.  Since the onset stress for shear thickening (up-pointing triangles in Fig. 3) lies above the yield line, the shear thickening and jamming regions are separated by a thin wedge of a shear thinning regime.  %This behavior is compressed to a very narrow range in $\phi$ for glass spheres due to the steepness of the yield line.  With more pronounced curvature of the yield line in the $\tau-\phi$ plane, as e.g. in cornstarch, it is therefore possible to move vertically at fixed $\phi$ from jamming to shear thickening via an intermediate shear thinning state, a sequence that has been termed "re-entrant jamming" by Fall $et al.$ \cite{FBHOB08}.  
The convergence of the yield stress and the boundaries between shear thickening and shear thinning suggests that $\phi_c$ is  a singular point.  We name this the {\it dynamic jamming point} because it is analogous to a static jammed state with a yield stress in the sense that the slope divergence implies the stress increases without an increase in shear rate,  i.~e., exhibits a jump in $\tau(\dot\gamma)$.   Furthermore, since the viscosity magnitude diverges in $\phi_c-\phi$ and the onset shear stress of the shear thickening regime is relatively fixed except for the influence of the yield stress the shear rate required for shear thickening extrapolates to zero at $\phi_c$. Thus, the discontinuous shear thickening limit corresponds to a jammed state.  We note, however, that the approach to zero onset shear rate cannot be measured all the way up to $\phi_c$ because the diverging yield stress suppresses shear thickening before $\phi_c$ is actually reached.  This extrapolation is made without reference to the yield stress measurements; the observation that this dynamic jamming point occurs on the jamming phase boundary confirms the correspondence.

The phase diagram in Fig.~\ref{fig:jam_phase} is valid for confined suspensions of non-attracting particles.  While details differ for cornstarch and glass sphere systems, both show qualitatively similar behavior so the overall delineation of jammed, shear thickening and shear thinning regions is robust.  The data shown were taken with smooth rheometer plates, but we also did experiments with roughened plates and saw no qualitative differences.  However, there is an important factor that can change the phase diagram as drawn in Fig.~\ref{fig:jam_phase}.  First, with attractive interactions  particles can chain across the system to produce a large yield stress at low packing fractions \cite{TPCSW01}. Such a yield stress can overwhelm any stress increase associated with shear thickening so that shear thickening is not observed, similar to the situation above $\phi_c$ in Fig.~\ref{fig:jam_phase}.   %Another factor is the magnitude of $\tau_{y,max}$; if it were much less than the stress at the upper end of the shear thickening regime, shear thickening could be observed all the way up to $\phi_c$ and possibly above $\phi_c$ if the viscosity remains finite.   Observations of a discontinuous shear thickening regime have been claimed  \cite{Laun94, BW96, BBS02, EW05, LDHH05, FHBOB08}, but in the absence of quantitative measurements of the yield stress or the viscosity slope remain speculative.   
The boundary conditions are important as well.  In the parallel plate setup, when the  experiment was done instead with extra fluid extending outside the plates the shear thickening curves were less steep, and a divergent scaling of the slope of the viscosity curve was not achieved.  This suggests that confinement (in this case by surface tension) is necessary to observe the approach to discontinuous shear thickening.  

%It has already been noticed that this change in boundary conditions also increases the shear rate at the onset of shear thickening \cite{FBHOB08}.  

Finally, we turn to the significance of the measured value for $\phi_c$.  Given the large uncertainty in determining an absolute packing fraction for starch samples due to water absorption from the atmosphere, due to the fact that dry powders generally pack less tightly than larger particles because of van der Waals forces, and due to the non-spherical polydisperse shapes of the starch particles (Fig. 2b), we focus here on the glass spheres.  We find near identical values for $\phi_c$ when the fluid is replaced by silicone oil or water, or when we double the particle size.  Thus, $\phi_c$ is not determined by fluid properties or particle diameter but rather seems to be related to an issue of geometric packing.  Our average value of $\phi_c = 0.573\pm0.007$ (including an absolute error of 0.005) for glass spheres is close to the value of random loose packing at 0.56 for glass spheres, corresponding to the packing fraction particles will reach when settled slowly in the limit of a density matched fluid \cite{OL90}.  This coincides with the onset of dilation, which is the lowest packing fraction where the packing must expand when sheared\cite{OL90}.  The value 0.56 is for large samples with density matching, and a positive correction of around 1\% would be expected both because of the finite size of our sample and the weight of the packing \cite{JSSSSA08}.  It is notable that shear thickening has long been associated with dilation  \cite{Hoffmann, OL90, LDHH05,FHBOB08}, mostly through normal force measurements, but a critical packing fraction has not been identified by such measurements. % On the other hand, our value for $\phi_c$ is also compatible with reported values $\phi_{max} \approx 0.58$ for the glass transition  in hard sphere colloids \cite{HRBWW00}, taken to be the point where diffusion of particles within the suspension effectively stops. The packing fraction $\phi_{max}$ is typically obtained by fitting the Krieger-Doherty relation $\eta = \eta_{fluid} (\phi_{max}-\phi)^{-2}$ to the zero-shear viscosities.  Prior results for shear thickening suspensions indicated $\phi_{max}$ values above the onset of discontinuous shear thickening, which would suggest $\phi_c$ occurs below the glass transition  \cite{EW05, note2}.  Given the uncertainty in our values of $\phi_c$ and the proximity of both the onset of dilation and the glass transition, we cannot rule out that there are two separate critical packing fractions.  

The existence of divergent scalings at a critical packing fraction is provocatively reminiscent of a 2nd order phase transition.   While the jamming point has been shown to have similarities to a critical point with some non-universal critical exponents \cite{OSLN03}, and some scalings compatible with a 2nd order phase transition \cite{OT07}, we are not aware of any such model that includes shear thickening.  %However, we cannot explore this further since we cannot accurately determine limiting critical exponents.

\section{Acknowledgements}

We thank T. Witten and S. Nagel for thoughtful discussions, and Jing Xu for help with the optical tweezer measurements. This work was supported by DARPA through Army grant W911NF-08-1-0209 and by the NSF MRSEC program under DMR-0820054.

\end{document}